\documentclass[jcp,aip,reprint,letterpaper,floatfix,onecolumn,nofootinbib,12pt]{revtex4-2}
\usepackage{amssymb,amsmath}
\usepackage[dvipsnames]{xcolor}
\usepackage{graphicx}
\usepackage{tabularx,booktabs}
\usepackage[labelfont=bf]{caption}
\usepackage[top=0.85in, bottom=0.85in,left=0.85in,right=0.85in]{geometry}
\usepackage[version=4]{mhchem}
\usepackage[colorlinks,urlcolor=Blue,citecolor=ForestGreen, linkcolor=Sepia]{hyperref}

\newcommand{\abs}[1]{\left|#1\right|}
\usepackage{multirow}
\usepackage{enumitem}

\begin{abstract}
    Efficient and reliable identification and optimization of transition state structures is a longstanding challenge in computational chemistry. Popular chain-of-states methods require hundreds if not thousands of \textit{ab initio} calculations to generate initial guesses for local quasi-Newton optimizers, with persistent risk of collapse to an alternative stationary point on the potential energy surface (PES). Here, we show that high-quality guess structures for transition state optimization can be obtained by constructing the geodesic path between reactant and product structures on the PES generated by machine learning potentials (MLPs). We present an algorithm for optimization of such geodesic paths, as well as the associated codebase. We demonstrate effectiveness of this approach using the recent eSEN-sm-cons MLP. On average, the highest-energy point along these MLP geodesics requires 30\% fewer quasi-Newton optimization steps to converge to the transition state compared to guesses from the fully \textit{ab initio} frozen string method. Our approach therefore completely eliminates the need for \textit{ab initio} calculations for generation of transition state guesses and considerably speeds up subsequent structural optimization. Geodesic construction on ML PES thus promises to be a useful approach for efficient computational elucidation of complex chemical reaction networks. 
\end{abstract}

\begin{document}
\title
{\large Locating Ab Initio Transition States via Approximate Geodesics on Machine Learned Potential Energy Surfaces}

\author{Diptarka Hait}
\author{Jan D. Estrada Pabón}
\author{Martin St\"ohr}
\author{Todd J. Mart{\'i}nez}
\email{todd.martinez@stanford.edu; toddjmartinez@gmail.com}

\affiliation{Department of Chemistry and The PULSE Institute, Stanford University, Stanford, California 94305, United States}
\affiliation{SLAC National Accelerator Laboratory, Menlo Park, California 94024, United States}

\maketitle

\section{Introduction}

\begin{figure}[hbt!]
    \begin{minipage}{0.48\linewidth}
        \includegraphics[width=\linewidth]{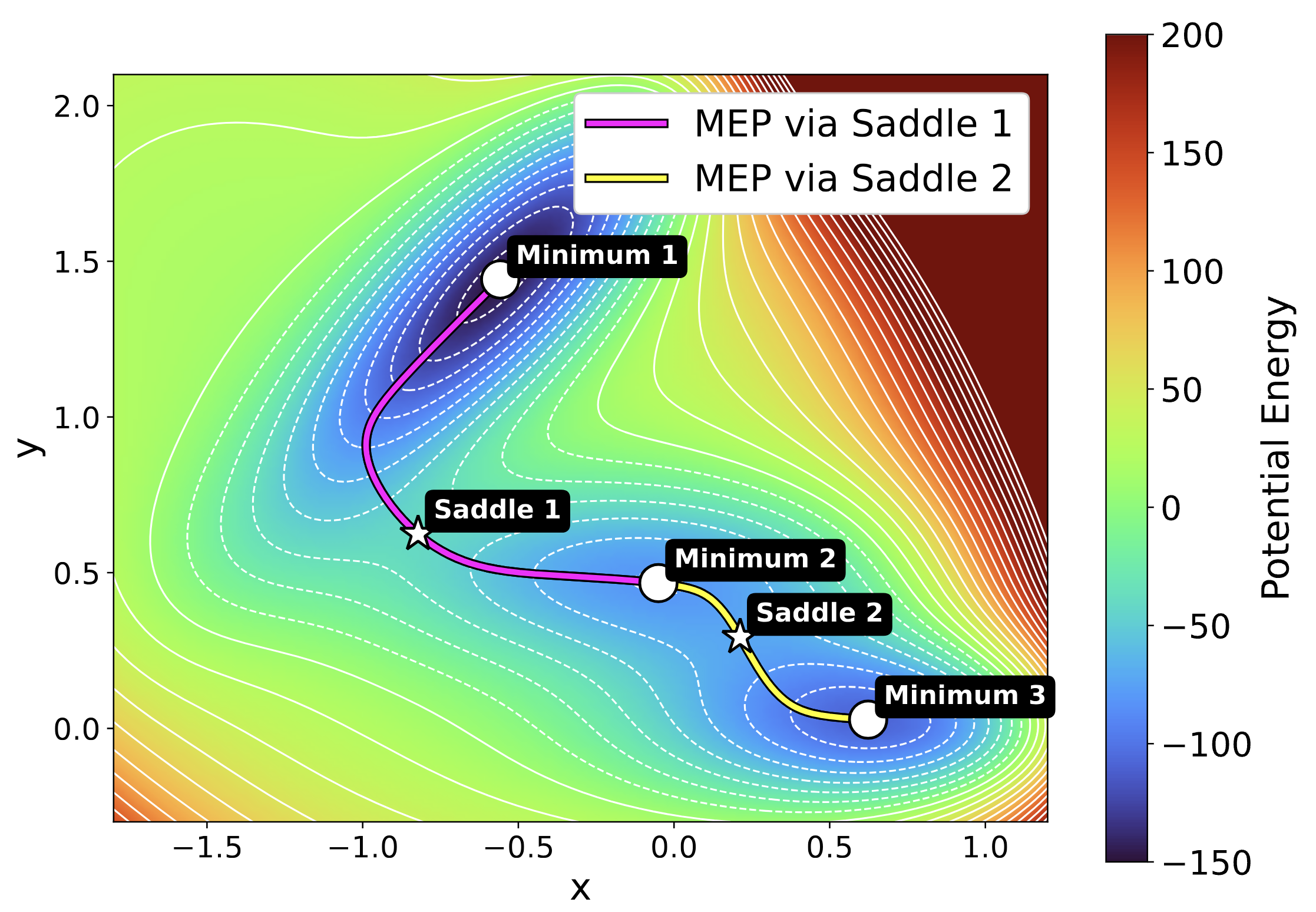}
    \end{minipage}
    \begin{minipage}{0.48\linewidth}
        \includegraphics[width=\linewidth]{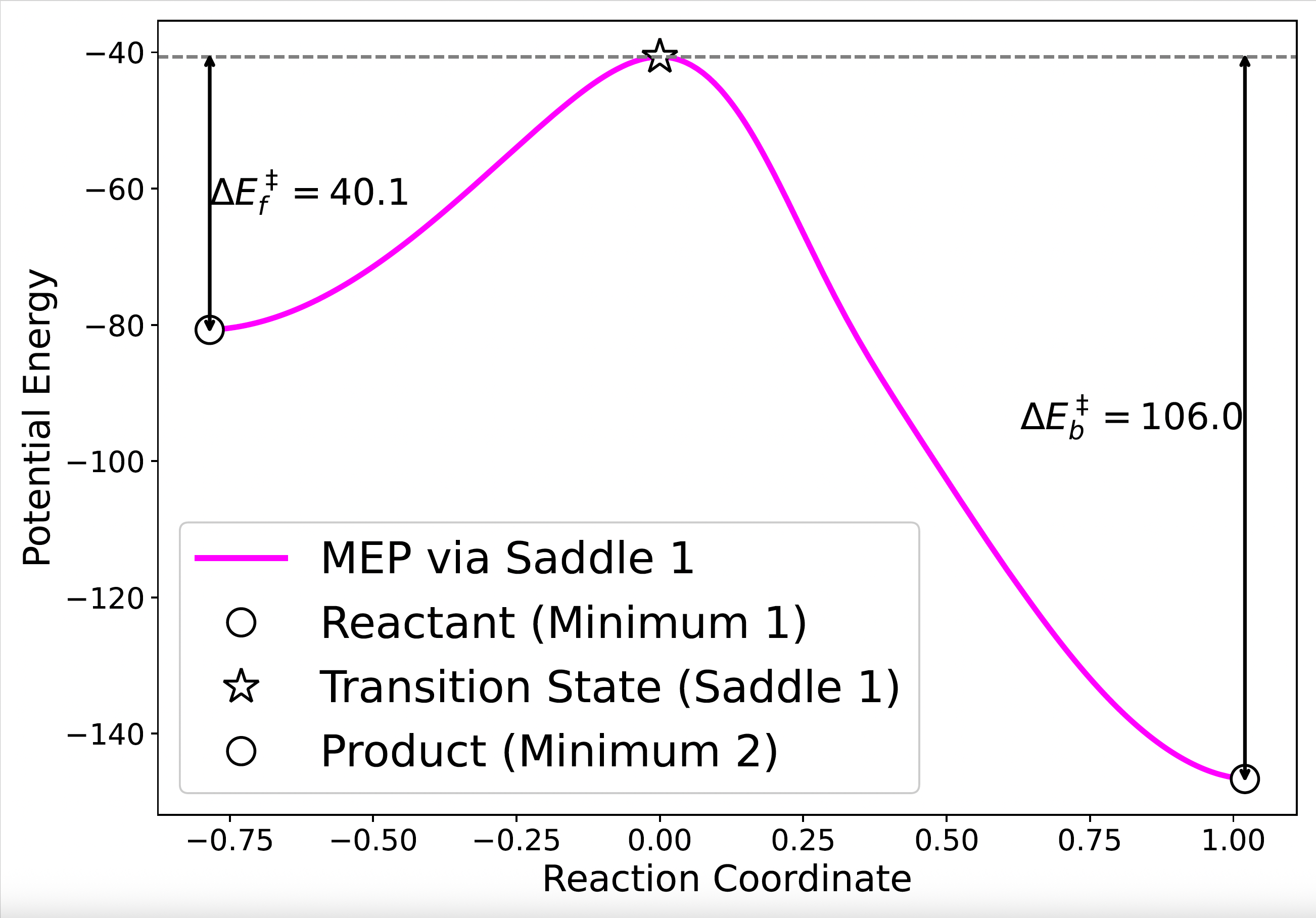}
    \end{minipage}
    \centering
    \caption{The model M{\"u}ller-Brown\cite{muller1979location} PES with stationary points and connecting MEPs highlighted (left) and the PES cross-section along the first elementary step connecting minimum energy points 1 and 2 via saddle 1 (right). Energy differences between the transition state and the minimum energy points $\Delta E^\ddagger$ for both the forward and backward processes are also shown for the right panel. }
    \label{fig:muller-brown}
\end{figure}

The feasibility of a chemical transformation is determined by the rate at which it occurs, as well as the kinetics of competing processes involving associated chemical species. Transition State Theory\cite{eyring1935activated,evans1935some} (TST) is the fundamental model for studying the kinetics of chemical reactions that occur on the electronic ground state.\cite{laidler1983development} In TST, reactions are decomposed into ``elementary'' steps between reactants and products, which are local minima on the Born-Oppenheimer potential energy surface (PES). For an elementary reaction, the minimum energy path\cite{johnston1961large} connecting the reactant and product passes through a single maximum of energy along the path. This point corresponds to a first-order saddle point on the overall PES (maximum along the MEP, minimum along all orthogonal directions) and is referred to as the transition state (TS). Fig.~\ref{fig:muller-brown} illustrates this concept using the model M{\"u}ller-Brown potential\cite{muller1979location}, showing how MEPs between minimum energy points involve passage through first-order saddle points.

In TST, the rate of an elementary reaction depends on the probability of the system reaching the TS. Under quasiequilibrium conditions, where the system can exchange energy with the environment at temperature $T$,
the reactant to product conversion rate is:
\begin{align}
    k\sim \dfrac{k_BT}{\hbar} \exp\left(-\dfrac{\Delta G^{\ddagger}}{k_BT}\right) \label{eq:tst}
\end{align}
where $\Delta G^{\ddagger}$ is the free energy difference between the TS and the reactant, and $k_B$ is the Boltzmann constant. In other words, the rate of an elementary chemical reaction is exponentially suppressed by the free energy difference $\Delta G^{\ddagger}$. While Eq.~\ref{eq:tst} applies only to elementary reactions, the rates of general chemical transformations can be modeled by decomposing the overall process into networks of elementary steps. Elementary steps therefore form the fundamental unit of rate theories and their identification is the central objective in the study of reaction mechanisms.

Although chemical reactions can often be quite slow, TSs typically have lifetimes on the order of a vibrational period (\textit{i.e.}, a few femtoseconds) and are difficult to observe experimentally.\cite{polanyi1995direct,manolopoulos1993transition,weichman2017feshbach,prozument2020photodissociation,ross2022jahn} The characterization of TSs is thus generally achieved through computation, typically using PES modeled with \textit{ab initio} calculations, usually density functional theory\cite{kohn1965self,mardirossian2017thirty,bursch2022best} (DFT). In fact, high-throughput generation of transition states is central to computational approaches to reaction discovery.\cite{wang2014discovering,suleimanov2015automated,dewyer2018methods,simm2018exploration,doentgen2018automated,woulfe2025chemical} However, the optimization of saddle point geometries is a more complex task than optimizing stable local minima.\cite{schlegel2011geometry} In particular, reaction coordinates are typically nonlinear functions of atomic positions that cannot straightforwardly be determined from reactant and product structures alone. 

The standard approach to locating TSs is to approximate the MEP using chain-of-states techniques,\cite{sheppard2008optimization} which discretize the path into a series of geometries (``nodes") whose positions are iteratively updated. 
Widely used examples include Nudged Elastic Band\cite{jonsson1998nudged,henkelman2000climbing} (NEB), String Method\cite{weinan2002string,burger2006quadratic} (SM), Growing String Method\cite{peters2004growing,zimmerman2013growing} (GSM) and Frozen String Method\cite{behn2011efficient,mallikarjun2012automated} (FSM). These methods are computationally demanding, as a large number of \textit{ab initio} force evaluations are required for node refinement.\cite{schlegel2011geometry} For instance, a canonical NEB calculation with $N$ nodes ($N\sim 20$ being typical) requires $N-2$ force evaluations at each iteration, with the endpoints being fixed. Since dozens of such iterations are often required for convergence, the total computational expense routinely runs to hundreds or thousands of \textit{ab initio} calculations for a single MEP.\cite{schlegel2011geometry,asgeirsson2021nudged}

In addition, the generation of initial node geometries is nontrivial. For instance, simple linear interpolation in Cartesian coordinates often creates unphysical structures with close contact between atoms, which can either result in incorrect high-energy reaction paths or require an enormous number of force evaluations to refine.\cite{rong2016efficient}
Moreover, chain-of-state methods only relax to locally optimal reaction paths without global exploration, which makes the discovery of MEPs critically dependent on the quality of the initial guess.
Consequently, better interpolation schemes like linear synchronous transit\cite{halgren1977synchronous} (LST), image dependent pair potential\cite{smidstrup2014improved} (IDPP) and Nebterpolator\cite{wang2016automated} have been developed. A particularly efficient approach for obtaining good initial node geometries is ``geodesic interpolation'',\cite{zhu2019geodesic} where interatomic distances are scaled with Morse potentials before interpolation to yield a physically reasonable path that aims to avoid close contacts between atoms. The term `geodesic' in this context was inspired by arguments from differential geometry about chemical reactions, which frame the reaction path as the shortest possible path (geodesic) on the PES for a specific choice of the metric that defines distance.\cite{tachibana1978differential,tachibana1979intrinsic} For the remainder of this work, we will refer to the interpolation approach from Ref.~\citenum{zhu2019geodesic} as `Morse-geodesic'. Despite the considerable improvements provided by these schemes, the path must subsequently still be converged through many costly \textit{ab initio} refinement steps. 

The highest-energy point obtained from approximate MEPs is subsequently used as an initial guess for quasi-Newton methods based on partitioned rational function optimization\cite{banerjee1985search,baker1986algorithm} (P-RFO) or the dimer method,\cite{henkelman1999dimer,kastner2008superlinearly} which directly optimize the TS geometry. The success of these local optimization methods strongly depends on the quality of the initial guess geometry. A poor guess can cause the optimization to converge to an undesired stationary point, such as an unstable conformer of the reactant/product or the TS of some other reaction. The development of computationally inexpensive methods for generation of high-quality guess geometries for TSs is thus of considerable interest. 

\begin{figure}[htb!]
    \centering
    \includegraphics[width=0.5\linewidth]{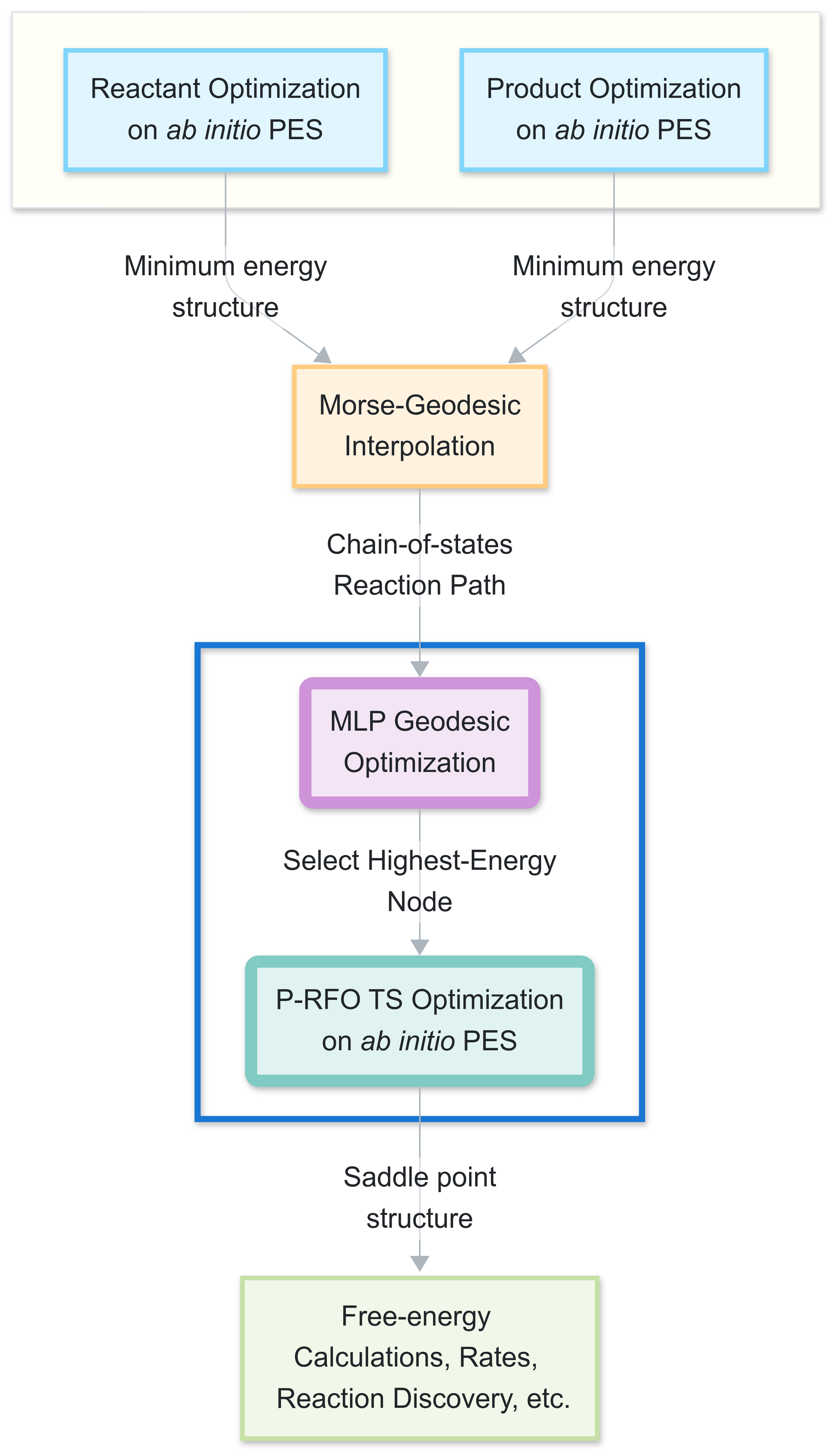}
    \caption{Scheme for transition state optimization with the protocol described in this work (with the steps most pertinent to this work bounded in blue). A Morse-geodesic path is initially interpolated between reactant and product structures that have been optimized on the \textit{ab initio} PES. This initial guess path is then optimized to a geodesic on the MLP. The highest-energy node geometry is subsequently optimized to a stationary TS geometry with P-RFO on the \textit{ab initio} PES, enabling the estimation of TST rates and other transition state properties.}
    \label{fig:tsopt}
\end{figure}

Machine Learning (ML) methods have attracted recent attention\cite{duan2023accurate,duan2025optimal,yuan2024analytical,marks2025efficient,choi2023prediction,li2025transition,anstine2025aimnet2,nakano2025high} for locating and characterizing TSs, with machine learned potentials\cite{behler2021four} (MLPs) offering computationally inexpensive approximations to the PES. In this work, we present an approach for constructing geodesic paths directly on an MLP. This approach reliably generates approximate  reaction path geometries without expensive \textit{ab initio} calculations. The highest-energy geometry along the MLP geodesic path is found to be a high-quality guess for direct TS optimization, with no need for further refinement via NEB or other chain-of-states methods.  We demonstrate this through direct P-RFO optimization of the TS on the \textit{ab initio} PES starting from the highest energy geometry on the MLP geodesic path. In fact, these MLP geodesic guess structures, generated without any \textit{ab initio} calculations, often converge to the TS in fewer P-RFO optimization steps than guess structures from the purely \textit{ab initio} FSM method. Crucially, our approach succeeds in generating viable TS guesses for challenging systems, such as ethene hydrogenation, where traditional FSM-based guesses struggle.\cite{marks2025efficient} Our protocol (as shown in Fig.~\ref{fig:tsopt}) therefore represents a highly cost-efficient route to locating TSs that would be useful for studying chemical kinetics and catalysis.

\section{Theory}
A geodesic is the shortest length path that connects two points on a Riemannian manifold, which thus depends on the metric tensor $\mathbf{g}$. For a chemical reaction the relevant (locally) Riemannian manifold is the PES, corresponding to the graph of the potential energy function, $U(\vec{R})$. In this case, Refs.~\citenum{tachibana1978differential} and \citenum{tachibana1979intrinsic} define $\mathbf{g}$ through $U(\vec{R})$ as:
\begin{align}
    \mathbf{g}(\vec{R}) = \biggl(\dfrac{\partial U}{\partial\vec{R}}\biggr)  \biggl(\dfrac{\partial U}{\partial\vec{R}}\biggr)^{\rm \!\!T}
    \label{eq:g-tensor}
\end{align}
(\textit{i.e.}, the outer product of the gradient with itself) in Cartesian coordinates.
The line element $ds$, which is the infinitesimal path length arising from an infinitesimal Cartesian coordinate displacement $d\vec{R}$, is then given by:
\begin{align}
    ds = \sqrt{\bigl(d \vec{R}\bigr)^{\rm \!T} \, \mathbf{g}(\vec{R}) \, d \vec{R}} = \sqrt{\Biggl( \biggl(\dfrac{\partial U}{\partial\vec{R}}\biggr)^{\rm \!\!T} d \vec{R}\Biggr)^{\!2}} =  \abs{\biggl(\dfrac{\partial U}{\partial\vec{R}}\biggr)^{\rm \!\!T} d\vec{R}\,}
\end{align}
Therefore, the path length $ \Delta s$ corresponding to evolution of the system from $\vec{R}_1\to\vec{R}_2$ is:
\begin{align}
    \Delta s = \displaystyle\int\limits_{\vec{R}_1}^{\vec{R}_2}  ds = \displaystyle\int\limits_{\vec{R}_1}^{\vec{R}_2} \abs{\biggl(\dfrac{\partial U}{\partial\vec{R}}\biggr)^{\rm \!\!T} d\vec{R}\,}
    \label{eq:Delta_s}
\end{align}
where the integral is evaluated over the specific path connecting the two points. 

The geodesic is the path connecting $\vec{R}_1\to\vec{R}_2$ while minimizing $\Delta s$. For an arbitrary path connecting $\vec{R}_1\to\vec{R}_2$, $\Delta s_{LB} = \bigl\vert U(\vec{R}_2)-U(\vec{R}_1)\bigr\vert$ is a lower bound to the true geodesic path length. In fact, if there exists a path connecting $\vec{R}_1\to\vec{R}_2$ for which $\biggl(\dfrac{\partial U}{\partial\vec{R}}\biggr)^{\rm \!\!T} d\vec{R}$ does not change sign (meaning there is a path $\vec{R}_1\to\vec{R}_2$ along which the energy is monotonic), then 
$\Delta s_{\rm LB}$ \emph{is} the geodesic path length.
For an elementary chemical reaction passing through a single first-order saddle point, we can split the true geodesic into two such monotonic parts: one connecting reactant to the TS and one from the product to the TS. The geodesic path length for elementary reactions is thus given by $\Delta s_{ECR} = \bigl\vert U(\vec{R}_{\mathrm{TS}})-U(\vec{R}_{\mathrm{react}})\bigr\vert + \bigl\vert U(\vec{R}_{\mathrm{TS}})-U(\vec{R}_{\mathrm{prod}})\bigr\vert$, \textit{i.e.}, the sum of the forward and backward reaction energy barriers. If there are multiple elementary channels connecting the reactant and product with different saddle point energies, the geodesic corresponds to the path through the lowest energy transition state.  

For an elementary chemical reaction, the MEP and geodesic are identical, making the optimization of reaction path lengths on the PES the central target of this work. 
For non-elementary reactions geodesics and MEPs need not necessarily coincide (as might be visually evident from Fig.~\ref{fig:muller-brown}) and the challenges of (multi-step) reaction path optimization are left for future work. We acknowledge that we do not cover the differential geometry of chemical reactions in detail in this work, and instead refer interested readers to Refs.~\citenum{zhu2019geodesic,tachibana1978differential,tachibana1979intrinsic} for further discussion. We also note that with our choice of metric (Eq.~\ref{eq:g-tensor}), identifying geodesics on the PES represents one realization (or interpretation) of the same mathematical problem as obtaining minimum action paths in the overdamped limit,~\cite{olender1997yet,vanden2008geometric,diaz2016comparison} or a variational calculus approach to finding the intrinsic/variational reaction coordinate.~\cite{crehuet2005reaction,aguilar2007applications,quapp2008chemical,birkholz2015path}
This is evident from the similarity between our path length (Eq.~\ref{eq:Delta_s}) and the ``scalar work"~\cite{olender1997yet,vanden2008geometric,diaz2016comparison} defined in the context of minimum action paths or the ``variational reaction energy"\cite{crehuet2005reaction,aguilar2007applications,quapp2008chemical,birkholz2015path} minimized in the variational reaction coordinate method.
In contrast to most of these works, we retain the integrand in Eq.~\eqref{eq:Delta_s} as the absolute value of gradient \textit{projected} onto the tangent along the path rather than the product of absolute values of the gradient and the path element. As we shall see below this choice leads to an extremization problem resembling the basic definition of the steepest descent path in terms of variational calculus as given in Ref.~\citenum{bofill2019calculus}, for example.
Note that, as pointed out previously,~\cite{birkholz2015path} both formulations are equivalent along the final steepest descent path, where the path tangent (and thus $d\vec{R}$) is always aligned parallel to the gradient $\biggl(\dfrac{\partial U}{\partial\vec{R}}\biggr)^{\rm \!\!T} d\vec{R}$.

In this work, we seek to construct a geodesic on the MLP that connects reactant and product structures previously optimized with a target \textit{ab initio} method. Consequently, the endpoints are not generally stationary points on the MLP surface and must be held fixed throughout the optimization. We discretize the connecting path with a series of $N$ intermediate nodes with Cartesian coordinates $\vec{R}_{1,2,\ldots N}$. The net result is a path of $N+2$ nodes including the end points, with $\vec{R}_0$ and $\vec{R}_{N+1}$ corresponding to the reactant and product. The total path length is $S=\displaystyle\sum\limits_{k=0}^N s_k$, where $s_k$ is the  segment length between nodes $\vec{R}_k$ and $\vec{R}_{k+1}$. 

We estimate $s_k$ via a \textit{locally quadratic} approximation to the energy profile along the \textit{linear Cartesian path} connecting nodes $\vec{R}_k$ and $\vec{R}_{k+1}$. Specifically, we assume:
\begin{align}
    U\left((1-\lambda)\vec{R}_k+\lambda \vec{R}_{k+1}\right) \approx U_k(\lambda)=a_k\lambda^2+b_k\lambda +c_k
\end{align}
where $\{a_k,b_k,c_k\}$ are obtained from solving $U_k(0)=U(\vec{R}_k)$, $U_k(1)=U(\vec{R}_{k+1})$, and $U_k\left(\dfrac{1}{2}\right)=U\left(\dfrac{\vec{R}_{k}+\vec{R}_{k+1}}{2}\right)$. The resulting $\{a_k,b_k,c_k\}$ are:
\begin{align}
    a_k &=  2U(\vec{R}_k)+2U(\vec{R}_{k+1})-4U\left(\dfrac{\vec{R}_{k}+\vec{R}_{k+1}}{2}\right)\\
    b_k &= -3U(\vec{R}_k)-U(\vec{R}_{k+1}) +4U\left(\dfrac{\vec{R}_{k}+\vec{R}_{k+1}}{2}\right)\\
    c_k &= U(\vec{R}_k)
\end{align}

The approximate path length $s_k$ is then:
\begin{align}
        s_k \approx  \displaystyle\int\limits_{0}^{1} \sqrt{\left(\dfrac{\partial U_k (\lambda)}{\partial\lambda }\right)^2 }\,d\lambda = \displaystyle\int\limits_{0}^{1} \abs{2a_k\lambda+b_k}\,d\lambda 
\end{align}   
We note that the absolute value function is not differentiable at zero. To improve stability in gradient-based optimization schemes, we therefore additionally introduce a regularization parameter $\epsilon^2$ into the integral for the path length to obtain an approximate regularized segment length:
\begin{align}
        s_k \approx \displaystyle\int\limits_{0}^{1} \sqrt{\left(\dfrac{\partial U_k(\lambda)}{\partial \lambda}\right)^2+\epsilon^2}\,d\lambda &= \displaystyle\int\limits_{0}^{1} \sqrt{(2a_k\lambda +b_k)^2+\epsilon^2}\,d\lambda\\
         &=\dfrac{1}{2a_k}\displaystyle\int\limits_{b_k}^{2a_k+b_k} \sqrt{x^2+\epsilon^2}\,dx \,\,\,\,\,\left(\text{where } x= 2a_k\lambda +b_k\right)\\
         &=\dfrac{1}{4a_k}\left[x\sqrt{x^2+\epsilon
      ^2}+\epsilon^2\ln \left(x+\sqrt{x^2+\epsilon^2}\right)\right]_{b_k}^{2a_k+b_k}
\end{align}
The resulting segment length $s_k$ is thus an analytical function of the energies and positions of nodes $k$ and $k+1$, and the Cartesian midpoint between them. We choose $\epsilon^2 = \sqrt[4]{\iota} \approx 1.2 \times 10^{-4}$ eV$^2$, where $\iota=2^{-52}$ is machine epsilon for double-precision numbers, \textit{i.e.} ``the smallest representable number such that $1 + \iota \neq 1$''. This relatively large value of $\epsilon^2$ enables use of the present approach with single-precision arithmetic as well, where machine epsilon is $2^{-23}\approx 1.2\times 10^{-7}$. If $\abs{a_k}<\epsilon^2$, the path is essentially linear, and $s_k\approx  \sqrt{b_k^2+\epsilon^2}$. 
We note that computing derivatives of $s_k$ with respect to node positions $\vec{R}_{k}$ and $\vec{R}_{k+1}$ is trivial if the energies, $U(\vec{R})$, and forces, $-\dfrac{\partial U}{\partial\vec{R}}$, can be readily obtained at the nodes and also at the Cartesian midpoint $\dfrac{\vec{R}_{k}+\vec{R}_{k+1}}{2}$. This is the case for MLPs, and is also achievable with \textit{ab initio} methods like DFT at a greater computational cost.

In principle, a sufficiently high density of nodes and $\epsilon \to 0 $ should yield the exact path length $S=\displaystyle\sum\limits_{k=0}^N s_k$, which can then be minimized through variation in node positions $\{\vec{R}_{k}\}$ to obtain the geodesic. In this limit, the movement of the nodes along the geodesic path itself does not affect the path length. This is not the case for a discretized path with \textit{relatively} few nodes and in practice movement tangential to the path needs to be controlled (similar to a NEB calculation). In particular, it is essential to guarantee that the TS region is adequately sampled and not have all nodes cluster around minima. To address these practical concerns, we modify the optimization in three ways:

Firstly, we extend the path length into a loss function $L=S +\beta \displaystyle\sum\limits_{k=0}^N \left(\dfrac{s_k}{\bar{s}}-1\right)^2$ for minimization, where the $\beta$ term adds a weak penalty that aims to ensure that individual $s_k$ do not significantly deviate from the mean segment length $\bar{s} = \dfrac{1}{N}\displaystyle\sum\limits_{k=0}^N s_k$. In other words, the penalty term ensures that each segment approximately covers the same scale of energy change. $\beta$ thus performs a similar role as spring constants in NEB, except it aims to achieve equal placement in energy and not position. We select $\beta$ = 1~kcal/mol in this work. 

Secondly, we remove the tangential component from the gradient of $S$. To this end, we define the unit tangent vector: 
\begin{align}
\hat{u}_{k\parallel} = \frac{\hat{u}_{k\rightarrow} + \hat{u}_{k\leftarrow}}{\bigl\Vert\hat{u}_{k\rightarrow}+\hat{u}_{k\leftarrow}\bigr\Vert} \quad\mbox{with}\quad \hat{u}_{k\rightarrow} = \dfrac{\vec{R}_{k+1}-\vec{R}_{k}}{\bigl\Vert\vec{R}_{k+1}-\vec{R}_{k}\bigr\Vert} \quad\mbox{and}\quad \hat{u}_{k\leftarrow}  = \dfrac{\vec{R}_{k}-\vec{R}_{k-1}}{\bigl\Vert\vec{R}_{k}-\vec{R}_{k-1}\bigr\Vert}\,\,.
\end{align}
We find that this definition improves performance over using $\vec{R}_{k+1}-\vec{R}_{k-1}$, as the nodes are generally not uniformly spaced in position but rather in energy. We then subtract the tangent component of the path length gradient $\left(\dfrac{\partial S}{\partial \vec{R}_k}\right)_{\parallel} = \Biggl(\hat{u}_{k\parallel}^{\rm T}\, \dfrac{\partial S}{\partial \vec{R}_k}\Biggr) \hat{u}_{k\parallel}$ from the gradient of the total loss $\dfrac{\partial L}{\partial \vec{R}_k}$.  Note that this choice means that only the tangent forces affecting the total path length $S$ are affected --- the penalty term can and should continue to push nodes along the tangent direction towards uniform segment lengths (and thus, uniform placement in energy).

Thirdly, we implement a `climbing node' option where the highest-energy node can be prevented from sliding down the path and is instead pushed uphill in energy. In this mode, the loss gradient associated with the highest-energy image $C$ is altered as follows:
\begin{align}
    \dfrac{\partial L}{\partial \vec{R}_C} \to \dfrac{\partial L}{\partial \vec{R}_C} -\left(\hat{u}_{C\parallel}^{\rm T}\, \dfrac{\partial L}{\partial \vec{R}_C}\right)\hat{u}_{C\parallel} -\alpha_{\mathrm{climb}}\left(\hat{u}_{C\parallel}^{\rm T}\, \dfrac{\partial U (\vec{R}_C)}{\partial \vec{R}_C}\right)\hat{u}_{C\parallel}
\end{align}
In other words, the only force experienced by the highest-energy node along the tangent $\hat{u}_{C\parallel}$ direction is $\alpha_{\mathrm{climb}}$ fraction of the energy gradient along that direction, pushing the node \textit{upwards} in energy. \textit{All} other tangent components (including from the penalty term) are removed. This approach is similar to the climbing image approach for NEB.\cite{henkelman2000climbing}

\begin{table}[htb]
\caption{Parameters used for the MLP geodesic path optimization in this work, as also described in the text. All listed parameters can be controlled by the user \textit{if} desired, though we caution against significant changes (e.g. setting $\alpha_{\mathrm{climb}} >1$) that may lead to unstable behavior.}
\label{tab:hyperparameters}
\centering
{
\setlength{\tabcolsep}{4pt}
\begin{tabularx}{\textwidth}{lXc}
\toprule
\multicolumn{1}{c}{\textbf{Parameter}}  & \multicolumn{1}{c}{\textbf{Description}} & \multicolumn{1}{c}{\textbf{Default Value}} \\
\midrule
\multicolumn{3}{c}{\textbf{Model Hyperparameters}} \\
\midrule
Segment length penalty ($\beta$) & Controls uniform placement in energy space. & 1~kcal/mol \\
Refinement interval ($\tau_{\text{refine}}$) & Number of iterations between successive refinement checks during the refinement stage. & 10 \\
Refinement cutoff (\texttt{cutoff}) & Percentage of segment length for triggering node insertion in the refinement stage. & 10\% \\
Tangent projection  & Enables projection of the component of the path length gradient that is tangent to the path. & True \\
Climbing image  & Enables the climbing image method in the refinement stage. & True \\
Climbing force scaling ($\alpha_{\mathrm{climb}}$) & Scaling factor for the climbing force component. & 0.5 \\
\midrule
\multicolumn{3}{c}{\textbf{Convergence Criterion}} \\
\midrule
FIRE stage 1 iterations  & Max iterations for initial relaxation stage. & 200 \\
FIRE stage 2 iterations  & Max iterations for subsequent refinement stage. & 500 \\
FIRE gradient tolerance  & Convergence tolerance for the $\Vert L\Vert_\infty$ gradient norm. & 0.01~eV/\AA \\
Convergence window  & Window size for checking convergence. & 20 iterations \\
Path length tolerance  & Convergence tolerance for the path length within a convergence window. & 0.25~kcal/mol \\
Barrier height tolerance  & Convergence tolerance for the forward/backward barrier height span within a convergence window. & 0.25~kcal/mol \\
\bottomrule
\end{tabularx}
}
\end{table}

With these choices to guide path length optimization with discretized nodes, we approximate the geodesic via a two-stage approach involving a initial path relaxation stage, and a subsequent careful refinement stage. 

\noindent \textbf{Path relaxation:} We obtain a Morse-geodesic initial guess path connecting the reactant and product structures and relax it on the MLP surface using the Fast Inertial Relaxation Engine\cite{bitzek2006structural} (FIRE) minimization algorithm as implemented in the Atomic Simulation Environment\cite{larsen2017atomic} (ASE). The climbing node feature is not active in this stage. This stage rapidly reduces $S$ from the initial Morse-geodesic guess. The Cartesian geometries along the path are translated/rotated via the Kabsch algorithm\cite{kabsch1976solution} after this stage concludes, in order to ensure each node is optimally aligned in space with its  neighbors. 

\noindent \textbf{Path refinement:} We subsequently do a more careful FIRE optimization with the climbing node option activated ($\alpha_{\text{climb}}=0.5$). Every $\tau_{\text{refine}}=10$ steps, we also assess the quality of the quadratic fit in each interval. This check is based on a user-supplied $\texttt{cutoff}$ percentage (here chosen to be $10\%$) of the corresponding segment length $s_k$ and proceeds via the following steps: 
\begin{enumerate}[label=\arabic*)]
    \item Align the geometries along the path via the Kabsch algorithm.
    \item For any interval where the quadratic fit has an internal local maximum (\textit{i.e} $a_k \le 0$ and $-\dfrac{b_k}{2a_k}\in (0,1)$), find the MLP energy of all such estimated quadratic energy maximum geometries  $ E_{\mathrm{MLP}}^\mathrm{quad} = U\left(\vec{R}_{k}-\dfrac{b_k}{2a_k}\left(\vec{R}_{k+1}-\vec{R}_{k}\right)\right)$.
    \item If $E_{\mathrm{MLP}}^\mathrm{quad} - \mathrm{max}\left(U(\vec{R}_k),U(\vec{R}_{k+1}),U\left(\dfrac{\vec{R}_{k}+\vec{R}_{k+1}}{2}\right)
    \right) > \mathtt{cutoff}\times s_k$ this geometry is an important high-energy point and is inserted into the path. 
    \item If, conversely, $E_{\mathrm{MLP}}^\mathrm{quad} - \mathrm{max}\left(U(\vec{R}_k),U(\vec{R}_{k+1}),U\left(\dfrac{\vec{R}_{k}+\vec{R}_{k+1}}{2}\right)\right) < -\mathtt{cutoff}\times s_k$ or  $E_{\mathrm{MLP}}^\mathrm{quad} < \mathrm{min}\left(U(\vec{R}_k),U(\vec{R}_{k+1}),U\left(\dfrac{\vec{R}_{k}+\vec{R}_{k+1}}{2}\right)\right)$, the quadratic fit is poor for this segment. This geometry is also inserted into the path in order to better sample this region.
    \item If any insertions are carried out, the entire path is re-aligned after all insertions using the Kabsch algorithm before the optimization continues.
\end{enumerate}
In our work, we found the FIRE algorithm to give consistent and reliable results for both the path relaxation and refinement stages. Suitable adaptation of (quasi-)Newton-type methods for geodesic path optimization with better convergence is subject to ongoing investigations.
Both the path relaxation and path refinement stages are considered converged when either the magnitude of the largest component of the loss gradient is smaller than 0.01~eV/\AA{} or the path stops changing. The latter criterion is satisfied when the variations in the total path length $S$, the forward energy barrier and the backward energy barrier are all less than 0.25~kcal/mol over 20 consecutive iterations. These relatively tight convergence criteria may not generally be necessary to obtain a high-quality TS guess and future work could explore the adequacy of less stringent stopping conditions. 

For elementary reactions, the optimized geodesic path should only have one local maximum of the MLP energy. This geometry should be used as the initial guess structure for full TS optimization using P-RFO or related methods on \textit{ab initio} PES. If the optimized path contains multiple energy maxima, this may indicate a non-elementary (multi-step) reaction or a poorly converged path. In such cases, the geometry of each local energy maximum along the path is a candidate for P-RFO refinement.

As discussed above, our method shares similarities with existing approaches aiming to find minimum action paths or the variational/intrinsic reaction coordinate. However, in contrast to previous works, we define the integrand in Eq.~\eqref{eq:Delta_s} as the absolute value of the gradient projected along the path, instead of the gradient magnitude. This allows for a significant simplification towards the analytical evaluation of the length of path segments, in conjunction with the assumption of locally quadratic energy profiles within each segment. We note that a similar approach of polynomial interpolation has been utilized for the variational reaction coordinate method.~\cite{birkholz2016path} Our approach also tries to achieve uniform spacing of nodes in energy, as opposed to most earlier works that aims for equal placement in position.\cite{olender1997yet,jonsson1998nudged}

\section{Computational Methods}
We compared the performance of our MLP geodesic approach for generating TS guess structures to the fully \textit{ab initio} FSM implementation\cite{behn2011efficient,mallikarjun2012automated} in the Q-Chem software package.\cite{epifanovsky2021software} The FSM calculations were performed with default settings, which includes LST interpolation and quasi-Newton BFGS node optimization. The highest-energy geometry from the MLP geodesic path (energy evaluated using the MLP) and the highest-energy node from the FSM path (evaluated at the \textit{ab initio} level) were used as initial guesses for transition state optimization with P-RFO. The P-RFO optimizations were performed using delocalized internal coordinates\cite{baker1996generation} unless specified otherwise, and were initialized with the exact nuclear Hessian at the starting geometry. The total number of P-RFO iterations is used as a measure of the quality of the transition state guess (fewer iterations indicating higher quality). We note that while we asked for 17 nodes for both MLP geodesic and FSM, additional nodes can be added by our node insertion procedure for MLP geodesic construction as well as by Q-Chem's FSM implementation. 

Our MLP geodesic constructions used the recent eSEN-sm-cons model developed by Meta,\cite{levine2025open,fu2025learning} which was trained on $\omega$B97M-V\cite{mardirossian2016omegab97m}/def2-TZVPD\cite{weigend2005balanced,rappoport2010property} data and is thus well positioned towards matching the performance of state-of-the-art for hybrid density functional theory for molecular systems.\cite{mardirossian2017thirty} We note that our codebase also allows for the use of MLPs such as UMA\cite{wood2025family}, MACE-OFF\cite{kovacs2025mace} and EGRET\cite{wagen2025egret} but this was not extensively explored as eSEN-sm-cons proved effective for the systems investigated in this work. 

We used B3LYP-D3(BJ)\cite{becke1993density,stephens1994ab,grimme2011effect}/def2-SVP\cite{weigend2005balanced} with the SG-3 grid\cite{dasgupta2017standard} for all \textit{ab initio} calculations in this work, including FSM, P-RFO TS optimization, as well as optimization of reactants and products. We note that B3LYP is known to systematically underestimate reaction barrier heights\cite{durant1996evaluation} due to delocalization error\cite{perdew1982density,cohen2008insights} and calculations involving transition states should preferably be carried out with other functionals with a greater proportion of exact exchange, such as modern range-separated hybrids.\cite{mardirossian2016omegab97m,bursch2022best} However, B3LYP generally provides reasonable TS geometries\cite{durant1996evaluation} and has been widely used in previous studies of TS structure generation.\cite{mallikarjun2012automated,asgeirsson2021nudged} We therefore selected  B3LYP-D3(BJ) for this work to demonstrate the general transferability of our approach to standard \textit{ab initio} PESs, as the functional is sufficiently distinct from the $\omega$B97M-V functional used to train the eSEN-sm-cons MLP.  

\section{Results}
The performance of our MLP geodesic construction approach was assessed on two datasets of main-group reactions that have been previously employed to characterize transition state finding methods. The first is a set of nine main-group reactions from Ref.~\citenum{mallikarjun2012automated} that has been used in the development of FSM.\cite{mallikarjun2012automated,marks2024incorporation,marks2025efficient} The second is a set of 121 main group reactions in Ref.~\citenum{asgeirsson2021nudged} for the development of the energy-weighted climbing-image NEB approach, compiled from combining earlier datasets.\cite{birkholz2015using,zimmerman2013reliable} Some reactions are shared between datasets, namely alanine dipeptide rearrangement, keto-enol tautomerization of acetaldehyde, the Diels-Alder addition of butadiene and ethene to form cyclohexene, as well as the dehydrogenation reactions of ethane, silane, and formaldehyde. We do not remove these repeated reactions from either dataset in our analysis. 

For both datasets, initial transition state geometries were obtained from previous work (Refs.~\citenum{marks2024incorporation} and \citenum{asgeirsson2021nudged}, respectively) and reoptimized with P-RFO at the B3LYP-D3(BJ)/def2-SVP level to provide stationary reference geometries. We obtained the reactant and product geometries for each reaction by first performing Intrinsic Reaction Coordinate (IRC) calculations\cite{fukui1970formulation} from the reference TS, followed by geometry optimization of the IRC endpoints. However, in certain cases, minimization of the IRC endpoints did not yield stable minima. In these instances, we first attempted to reoptimize the corresponding reactant and product geometries provided in Refs.~\citenum{marks2024incorporation} and \citenum{asgeirsson2021nudged} with B3LYP-D3(BJ)/def2-SVP. However, for a small number of cases, using these previously reported structures introduced unnecessary rotations/distortions into the reaction path. For these cases, the corresponding endpoint geometries were found by either permuting identical atoms in the original structures, or manually displacing the original unstable IRC endpoint along its imaginary frequency mode(s) and subsequent reoptimization. All the stationary point geometries obtained in this work and additional information regarding their provenance is provided in the supporting information\citenum{hait_2025_16255159}. 
\begin{table}[htb]
\caption{Performance of P-RFO initialized with TS guess geometries from MLP geodesic, FSM, and Morse-geodesic. Ethane dehydrogenation fails to converge after 200 P-RFO iterations with delocalized internal coordinates for the FSM guess (and converges to an incorrect structure when starting from the highest-energy Morse-geodesic geometry), and therefore iteration counts for optimization with Cartesian coordinates are provided in parentheses. The highest-energy Morse-geodesic geometry for the Ireland-Claisen rearrangement fails to converge after 200 P-RFO iterations with delocalized internal coordinates as well.}
\label{tab:fsmdev}
\centering
\begin{tabular}{l c c c }
\toprule
 & \multicolumn{3}{c}{\textbf{P-RFO iterations starting from}}  \\
\textbf{Reaction} & \textbf{ MLP geodesic} & \textbf{FSM}  & \textbf{Morse-geodesic}\\
  & \textbf{guess} & \textbf{guess} & \textbf{maximum} \\
\midrule
\ce{H2CO -> H2 + CO} & 3 & 8 & 27  \\
\ce{SiH4 -> SiH2 + H2} & 3 & 3 & 5  \\
\ce{CH3CHO -> H2C=CHOH} & 3& 4 & 7\\
2,4-hexadiene \ce{->} 3,4-dimethylcyclobutene & 10 & 9  & 10 \\
Alanine dipeptide rearrangement & 58 & 56 & 46  \\
\parbox[t]{3in}{\raggedright Ireland–Claisen rearrangement between silyl ketene acetal and silyl ester} & \multirow{2}{*}{59} & \multirow{2}{*}{69} & \multirow{2}{*}{fails} \\
\ce{CH3CH3 -> H2C=CH2 + H2}  & 4 (4) & fails (24) & fails (23)  \\
bicyclobutane \ce{-> H2C=CH-CH=CH2}  & 4 & 39 & 31   \\
\ce{H2C=CH-CH=CH2 + H2C=CH2 ->} cyclohexene  & 6 & 22 & 26  \\
\bottomrule
\end{tabular}
\end{table}
\subsection{FSM development dataset}
Table \ref{tab:fsmdev} reports the number of P-RFO steps required to optimize the TS for the FSM development dataset, starting from initial guesses generated by our MLP geodesic approach and by FSM. The results indicate that P-RFO starting from our MLP geodesic generated guess leads to the reference TS in all cases. Four of the reactions (formaldehyde dehydrogenation, silane dehydrogenation, keto-enol tautomerization of acetaldehyde, and electrocyclic [2+2] ring closure of 2,4-hexadiene) converge readily with either method, requiring ten or less steps at the B3LYP-D3/def2-SVP level of theory.  This indicates that our MLP geodesic approach can generate structures that are of high quality and transferable to PES other than the $\omega$B97M-V/def2-TZVPD method used to train the MLP. We however note that simply picking the geometry with the highest B3LYP-D3/def2-SVP energy out of the initial Morse-geodesic interpolation is a sufficiently reasonable guess for three of these reactions, only being significantly inferior for the dehydrogenation of formaldehyde. 

Conversely, both MLP geodesic and FSM guesses require a fairly large number of steps for the two largest systems in the test set, namely rearrangement of alanine dipeptide and the Ireland-Claisen rearrangement of silyl ketene acetal into a silyl ester. Both require more than $50$ steps to optimize to the TS structure on relatively flat PES. 
We also note that the use of the highest-energy Morse-geodesic geometry surprisingly led to slightly faster convergence to the transition state for alanine dipeptide, while completely failing to converge in 200 iterations for the Ireland-Claisen reaction. 

The third set of systems are ethane dehydrogenation, rearrangement of bicyclobutane to 1,3-butadiene, and the Diels-Alder reaction between 1-3-butadiene and ethene. The MLP geodesic guess appears to be particularly advantageous for these three cases, yielding TS guesses that take almost an order of magnitude fewer P-RFO steps as compared to FSM or the highest-energy Morse-geodesic node. Ethane dehydrogenation is a particularly challenging case\cite{marks2025efficient} with the other two protocols for generating TS guess geometries failing to converge in over 200 steps with delocalized internal coordinates. Optimization with Cartesian coordinates fares somewhat better in this case, yielding the correct TS in 24 steps from the FSM guess, 23 steps from the highest-energy Morse-geodesic geometry, and 4 steps for the MLP geodesic guess. 

The bicyclobutane rearrangement reaction highlights another general challenge for TS optimizations, as the saddle point geometry obtained with spin-restricted B3LYP/def2-SVP is unstable against spin-symmetry breaking.\cite{seeger1977self,bauernschmitt1996stability} The resulting spin-contamination is not excessive ($\langle S^2 \rangle$ = 0.24 at the unrestricted B3LYP/def2-SVP optimized TS geometry with the same level of theory), and the spin-polarized transition state geometry is quite similar to the spin-restricted transition state geometry. The 
MLP geodesic guess consequently takes 5 P-RFO iterations to converge to the TS optimized at the spin-unrestricted level while the FSM guess requires 53 iterations (compared to 4 and 39, respectively, for the fully spin-restricted calculations). However, this case highlights a general challenge for high-energy gas phase processes, where ``elementary" steps detected with spin-restricted methods may be energetically suboptimal compared to pathways involving spin-polarization, including multistep processes involving labile open-shell species. These processes are challenging to model with even unrestricted Kohn-Sham DFT due to spin-contamination affecting the PES.\cite{yamaguchi1988spin,hratchian2013communication,hait2019well} MLPs, which are essentially advanced force-fields without explicit knowledge of electronic structure, are also likely to be challenged by such cases, even if spin-unrestricted DFT data is present in the training set (as is the case for eSEN-sm-cons). 

Overall, the analysis of these nine reaction clearly reveals that the MLP geodesic construction yields initial guess geometries that are of FSM quality or better, as judged by the number of P-RFO steps taken for convergence to the stationary TS structure. The MLP geodesic guesses are obtained from reaction endpoint geometries without additional \textit{ab initio} calculations, and are thus a computationally inexpensive approach for reaction path generation. This is in contrast to FSM, which typically requires $\sim$50 \textit{ab initio} gradient calls per 17 node calculation to obtain the full frozen string pathway. 

\subsection{Energy-weighted climbing-image NEB dataset}
We next evaluated the MLP geodesic approach on the dataset from Ref.~\citenum{asgeirsson2021nudged}, which originally contained 121 main-group reactions with stationary point geometries optimized with B3LYP-D3(BJ)/def2-SVP. We identified 27 cases where spin-symmetry breaking occurred in the supplied TS structures, often leading to significant change in geometry upon P-RFO reoptimization with spin unrestricted B3LYP-D3(BJ)/def2-SVP calculations. In addition, there were five other reactions where no spin-symmetry breaking was observed, but the B3LYP-D3(BJ)/def2-SVP P-RFO optimization required more than 15 iterations even though the starting geometries were reportedly optimized at the same level of theory. Three of these reactions involved proton transfer in the taxadiene cation\cite{zimmerman2013reliable}, for which the saddle point structures provided in Ref.~\citenum{asgeirsson2021nudged} appear to have been erroneously optimized under the assumption that the system is a neutral radical doublet. A pair of reactions also appear to be duplicated, having the same transition state, reactant and product structures.

We therefore removed all 32 cases where there was either spin-symmetry breaking or more than 15 P-RFO iterations for TS re-optimization starting from the provided saddle point geometry. We also removed a copy of the duplicated reaction. This left us with a set of 88 reactions for which we analyzed the performance of MLP geodesic and FSM. On the \textit{ab initio} PES, the maximum number of P-RFO steps required to reoptimize the provided saddle point geometries for these 88 processes was 7, with the majority requiring just 2 steps. Consequently, we have relatively high confidence in the quality of this curated set of 88 unique reactions. 

\begin{figure}[htb!]
    \begin{minipage}{0.48\linewidth}
        \includegraphics[width=\linewidth]{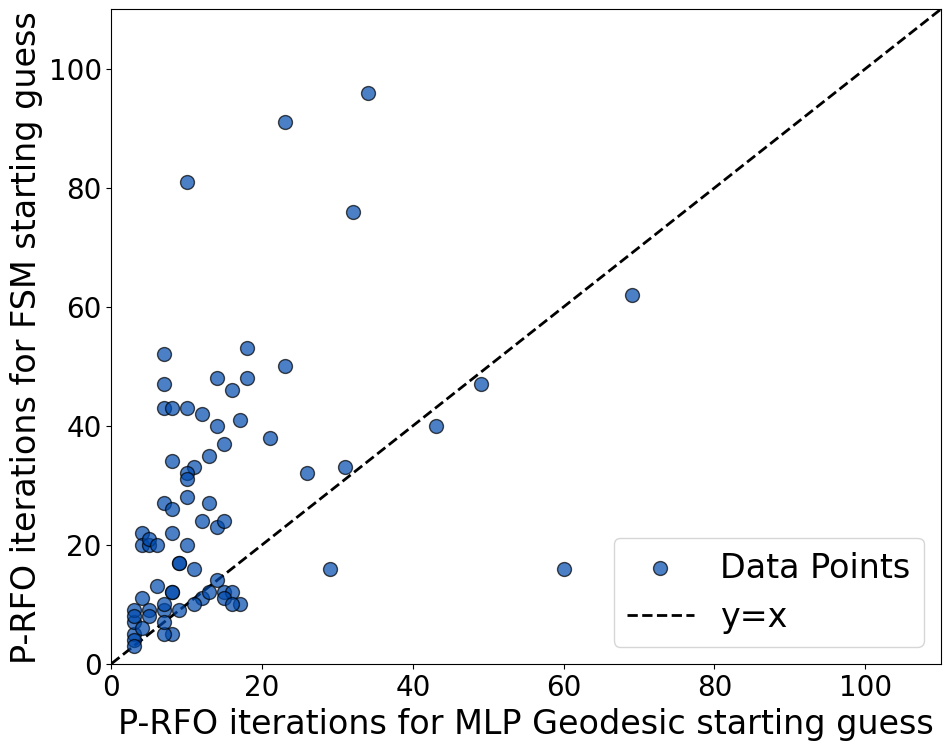}
    \end{minipage}
    \centering
    \caption{Comparison of the number of P-RFO iterations required for converging the FSM and MLP geodesic starting guesses to the transition state structure for 76 reactions in the dataset from Ref.~\citenum{asgeirsson2021nudged}. Each point corresponds to a separate reaction.}
    \label{fig:hannescomp}
\end{figure}

Of these 88 reactions, we find that there are 76 for which P-RFO optimizations with delocalized internal coordinates converges to the reference saddle point geometries when starting from the initial guess structures generated from either the MLP geodesic or FSM. As shown in Fig.~\ref{fig:hannescomp}, the P-RFO iteration counts for the two starting guesses are often similar, or the FSM guess requires notably more iterations (as indicated by the higher density of points above the $y$=$x$ line). However, there are 15 reactions for which the FSM generated guess leads to faster convergence, with the most significant example being the S\textsubscript{N}2 addition of acetate to an epoxide. The TS guess generated from the MLP geodesic requires 60 iterations to converge to the \textit{ab initio} TS for this reaction, while the FSM guess requires 16. This behavior is not too surprising, as it involves charged species and we have not explicitly provided charge information to the eSEN-sm-cons MLP.

\begin{figure}[htb!]
    \begin{minipage}{0.48\linewidth}
        \includegraphics[width=\linewidth]{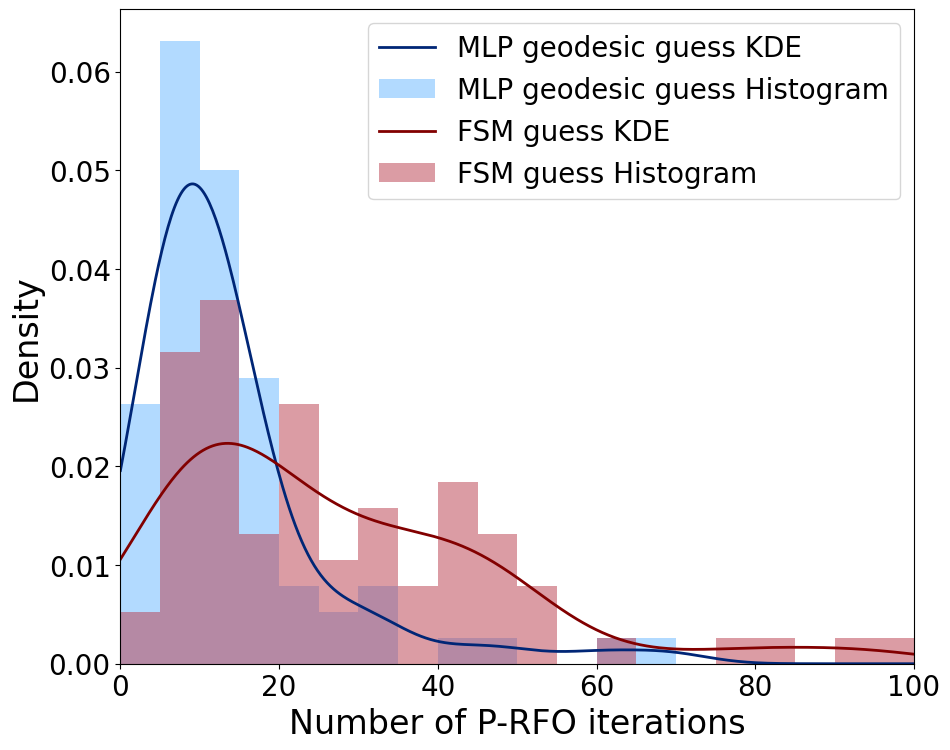}
    \end{minipage}
        \begin{minipage}{0.48\linewidth}
        \includegraphics[width=\linewidth]{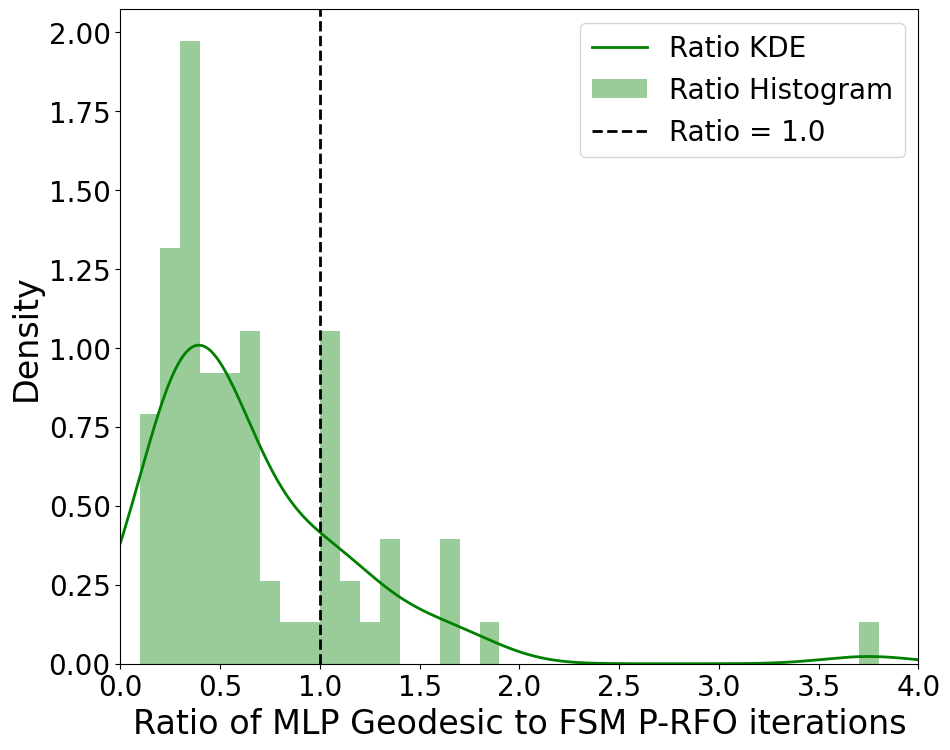}
    \end{minipage}
    \centering
    \caption{Histograms of the number of P-RFO iterations required for converging the FSM and MLP geodesic starting guesses to the transition state structure for 76 reactions in the dataset from Ref.~\citenum{asgeirsson2021nudged} (left) and the ratio of the iterations needed from MLP geodesic generated guesses to FSM generated guesses (right). A ratio less than 1 indicates a more efficient MLP geodesic guess. The corresponding kernel density estimates (KDE) of the associated probability distribution is also shown. }
    \label{fig:hannescomphist}
\end{figure}

To further analyze the performance, Fig.~\ref{fig:hannescomphist} shows the histograms of the P-RFO iteration counts starting from the MLP geodesic and FSM guesses for these 76 reactions, as well as the ratio between the iterations needed for each system.
The histograms of the individual P-RFO iterations for both sets of guesses (left panel) show that the MLP geodesic approach leads to faster and more consistent TS convergence, as the distribution of iteration counts is centered around a smaller modal value and is narrower. The histogram of the ratio of P-RFO iterations between the guesses generated by two methods also show MLP geodesic to be generally faster, with 38 reactions requiring less than half the number of steps as FSM. The average ratio is $0.66$ with a standard deviation of $0.54$. This further confirms that, from the P-RFO perspective, MLP geodesic construction generates guess structures that are on average closer to the true transition state.

\begin{figure}[htb!]
    \begin{minipage}[b]{0.48\linewidth}
    \includegraphics[width=\linewidth]{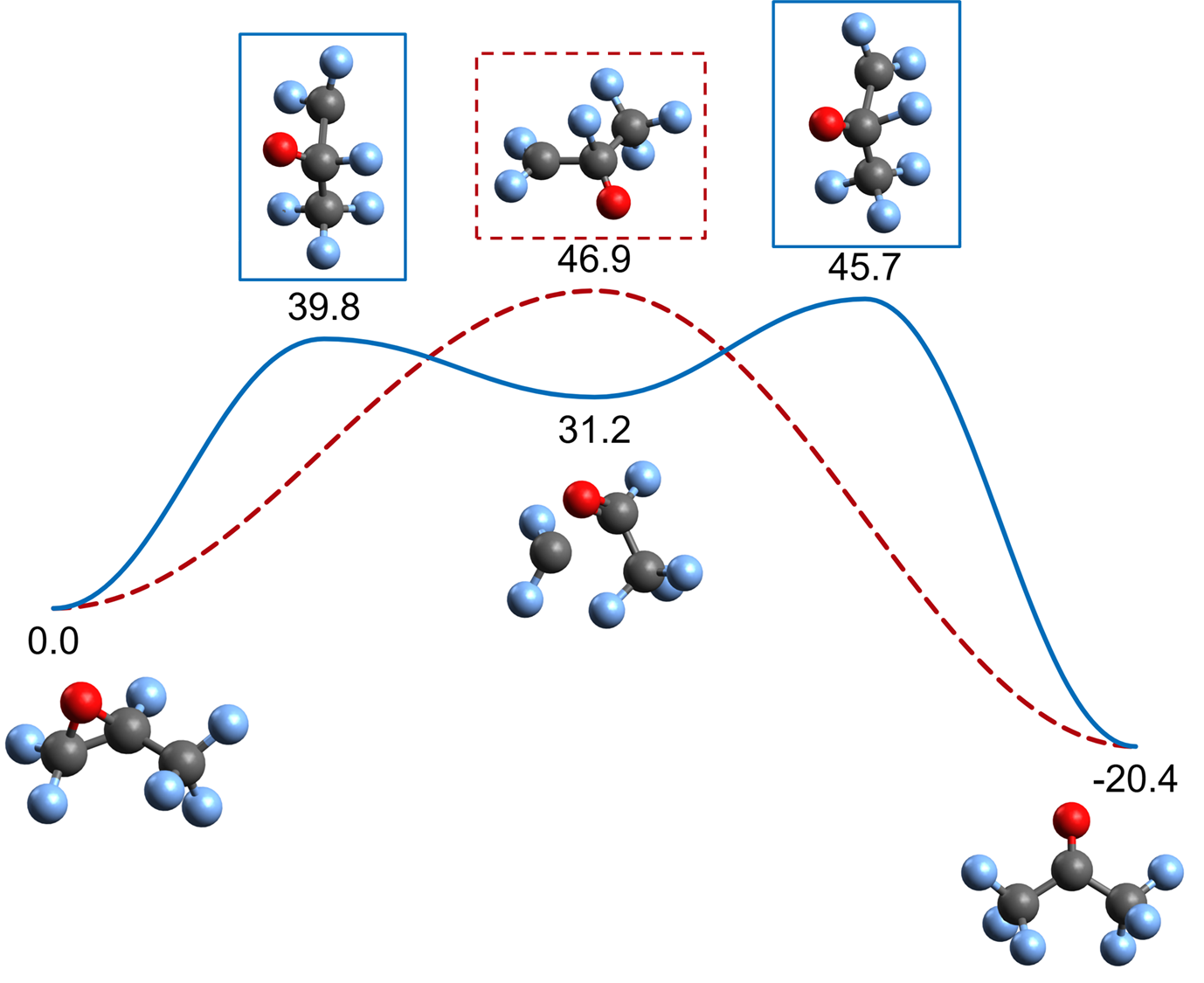}
    \centering
    \end{minipage}
    \begin{minipage}[b]{0.48\linewidth}
    \includegraphics[width=\linewidth]{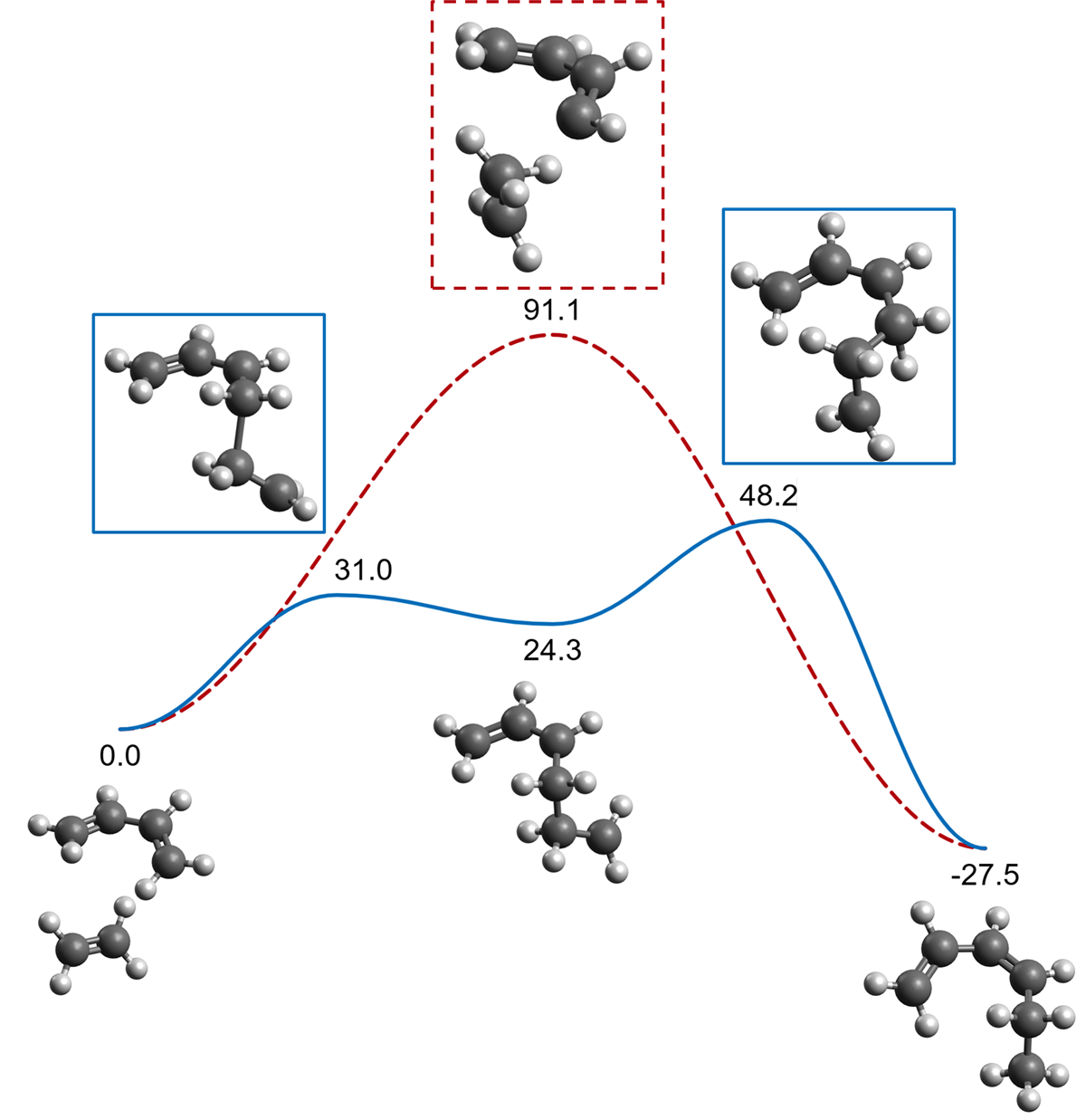}
    \centering
    \end{minipage}
    \caption{Two reactions in the dataset from Ref.~\citenum{asgeirsson2021nudged} for which MLP geodesic construction identifies a lower energy two-step pathway (blue solid line) over a higher energy elementary process (red dotted line): Rearrangement of perfluoropropylene oxide to perfluoroacetone (left panel) and addition of ethene to butadiene to form 1,3-hexadiene (right). The corresponding transition state structures (bordered by the line-style corresponding to the reaction mechanism) and the intermediate for the two-step path are also shown, as are energies of all stationary points (in kcal/mol, relative to the reactant). Energies are not to scale.}
    \label{fig:twosteps}
\end{figure}

We next consider the remaining 12 reactions for which either the MLP geodesic or FSM guesses do not converge to the reoptimized reference TS with delocalized internal coordinates. For 7 reactions (listed in the supporting information), the MLP geodesic guess converges to the reference TS while FSM either fails to converge or converges to an incorrect structure. The MLP geodesic guess does not converge to the reoptimized reference TS for the remaining five. In one case, the MLP geodesic guess leads to a demonstrably incorrect result. For the reaction between 2-butene-1-ol and \ce{SOCl2} leading to HCl and \ce{CH3CH=CHCH2OSOCl}, P-RFO optimization of the MLP generated TS guess structure leads to a higher energy unstable conformer of the reactant that is 15~kcal/mol below the reference TS structure. In the remaining four cases however, MLP geodesic discovers a more energetically favorable reaction pathway than the elementary step defined by the reference TS structure. One reaction (boron terminus hydrogen exchange between \ce{NH3BH3} and \ce{NH2BH2}) remains elementary, but optimization of the MLP geodesic generated guess leads to a lower energy TS (by 0.5~kcal/mol) than the reference, with this TS also leading to the correct endpoints upon an IRC calculation. 

The remaining three reactions have multistep pathways for which the highest-energy TS structure is below the reference TS at the B3LYP/def2-SVP level of theory. The left panel of Fig.~\ref{fig:twosteps} shows the competing pathways for the reorganization of perfluoropropylene oxide into perfluoroacetone. The reference TS for the elementary step is 46.9~kcal/mol above the reactant. However, the two-step process involving dissociation of a \ce{CF2} carbene moiety and subsequent recombination plus fluorine transfer to produce perfluoroacetone is \textit{slightly} energetically favored, with the TS for the dissociation step, the dissociated intermediate and the second TS for the recombination lying 39.8~kcal/mol, 31.2~kcal/mol, and 45.7~kcal/mol above the reactant at this level of theory. We note that the energy difference between the two pathways is ultimately quite small, and FSM finds the single-step path. In this regard, it is worth noting that CCSD(T)\cite{raghavachari1989fifth}/aug-cc-pVTZ\cite{dunning1989gaussian,kendall1992electron} on the B3LYP-D3(BJ)/def2-SVP optimized stationary structures gave very similar relative energies (elementary TS at 47.4~kcal, the two TSs of the two-step process at 39.2~kcal/mol and 46.6~kcal/mol, respectively).

The other two multi-step reactions are between butadiene and ethene, with one forming 1,3-hexadiene, and the other being a molecular reorganization leading back to butadiene and ethene. The reference TS structures from Ref.~\citenum{asgeirsson2021nudged} did not exhibit spin-symmetry breaking, but were quite high in energy (suggesting barriers of 91~kcal/mol and 124~kcal/mol, respectively, relative to the reactant). The MLP geodesic generated structures were significantly lower in energy, and exhibited spin-symmetry breaking. P-RFO optimization of these saddle point structures with spin-unrestricted B3LYP-D3(BJ)/def2-SVP, and subsequent IRC calculations revealed more energetically favorable multistep pathways involving biradicaloid intermediates. The right panel of Fig.~\ref{fig:twosteps} shows the pathways leading to 1,3-hexadiene formation. The two-step pathway involves $\eta^1$ addition of ethene to the terminal carbon of butadiene to form a biradicaloid adduct, which subsequently undergoes intramolecular hydrogen transfer to form 1,3-hexadiene, resulting in a pathway where the highest-energy TS (the hydrogen transfer one) is only 48.2~kcal/mol above the reactants. This pathway however involves biradicaloids that are greatly destabilized by spin-restricted calculations, highlighting a potential strength of the MLP geodesic approach when a MLP trained on spin-unrestricted data is utilized. The multistep biradicaloid pathway for the molecular reorganization leading back to butadiene and ethene is provided in the supporting information. 

Overall, we find that the MLP geodesic approach is quite effective at locating transition states for the well-behaved reactions from Ref.~\citenum{asgeirsson2021nudged}, with only one explicit failure. Indeed, the guess structures generated from the MLP geodesic typically lead to faster P-RFO convergence to the \textit{ab initio} saddle point than FSM. 

\section{Conclusions}

This work demonstrates that constructing geodesics on a state-of-the-art MLP (as exemplified by eSEN-sm-cons) is a very efficient strategy for generating high-quality TS guess structures, requiring no \textit{ab initio} calculations. The resulting geometries appear to be better than or comparable to those generated from FSM, requiring $\sim$30\% less P-RFO iterations (and thus \textit{ab initio} force evaluations) to converge to the final TS structure. Our approach is therefore a promising route towards reliable optimization of TSs with a minimal number of \textit{ab initio} calculations. We anticipate that geodesic construction on MLPs will soon become widely used for finding reaction paths and obtaining TS guess structures. The efficiency of such approaches would make them particularly amenable for finding the lowest energy reaction path between different conformers of reactants and products, which is a challenge in computational molecular catalysis. Although not explored in this work, generalization to periodic systems for heterogeneous catalysis applications is also straightforward. 

However, the promise of this approach comes with some caveats. At present, it is assumed that the MLP energy/force calculations are essentially free compared to the final TS refinement step on the \textit{ab initio} PES. This is a reasonable assumption that is likely to hold true in the near future due to developments in both software and hardware towards more efficient neural network calculations. Nonetheless, obtaining MLP geodesics typically involves hundreds of iterations over $\sim$40 nodes and Cartesian midpoints, although batched evaluation of energies/forces on GPUs significantly accelerates the process. It is quite likely that substantial MLP geodesic construction speedups are possible through more selective updating of nodes along the geodesic, better path optimization algorithms than FIRE, more carefully chosen model hyperparameters, and the potential use of internal coordinates instead of Cartesian coordinates. MLP Hessians can also help to speed up calculations by, \textit{e.g.}, providing normal mode information to identify the true reaction coordinate at a given node. 
In this context, future work will also explore the connections between constructing geodesic paths and identifying MAPs in more detail, which can further aid the methodological development of both approaches.
Finally, the initial \textit{ab initio} Hessian calculation required by P-RFO is a major computational bottleneck in the overall workflow and replacing it with an MLP Hessian could offer substantial cost savings.\cite{yuan2024analytical} 

One major limitation of this work is the focus on elementary reactions. Although our approach can lead to the discovery of multistep reactions (as shown in Fig.~\ref{fig:twosteps}), it does not do so automatically, and instead involves manual inspection of intermediate IRC calculations. A fully automatic procedure that leverages the MLP to identify distinct elementary steps and construct independent geodesic pathways would be desirable for high-throughput reaction discovery. One possible approach in this regard may be to extend the path refinement procedure to sample local minima along the path, and minimize/freeze any sufficiently deep minimum energy geometries. However, this will only be successful if the MLP is sufficiently accurate. In practice, MLPs could be qualitatively inaccurate for out-of-equilibrium high-energy structures not sampled during the training process. In fact, we observed an example during this study, while studying the hydrogen exchange between formaldehyde and vinyl alcohol with the EGRET-1T MLP. In this case, a NEB calculation led to the discovery of a rather unphysical structure (provided in the supporting information), which EGRET-1T predicted to be $\sim$190~kcal/mol below the reactants/products in energy. The eSEN-sm-cons MLP appears to be robust against such spurious behavior for the systems we have tested, but this does not guarantee good performance in all regimes (especially when charges/spins/long-range interactions become significant). The possibility of such structural collapse is a strong reason for the use of geodesic construction over other chain-of-state methods, as the minimization of total path length is a robust guardrail against nodes collapsing into spurious deep wells. However, this feature also makes discovery of multistep pathways more challenging. Hybrid approaches that adaptively switch between the MLP and \textit{ab initio} PES (or which integrate information from both) could be particularly useful in avoiding unphysical behavior while still leading to efficient elucidation of the reaction path. Work along these directions is currently in progress.

\section*{Acknowledgments} 
The authors thank Dr. Jutta Rogal for helpful discussions about minimum action paths. 
This research was financially supported by the Office of Naval Research (N00014-21-1-2151). D.H. is a Stanford Science Fellow. J.D.E.P acknowledges support by the National Science Foundation Graduate Research Fellowship, under Grant No. DGE-2146755. M.S. acknowledges financial support from the Deutsche Forschungsgemeinschaft (DFG, German Research Foundation) -- 534068594. This work used computational resources of the National Energy Research Scientific Computing Center (NERSC), a U.S. Department of Energy Office of Science User Facility located at Lawrence Berkeley National Laboratory, operated under Contract No. DE-AC02-05CH11231 using NERSC award BES-ERCAP0028744.

\section*{Supporting Information}
\noindent \textbf{Zenodo repository\cite{hait_2025_16255159}:} Code, Q-Chem output files. 

\noindent \textbf{PDF:} Additional information about specific reactions (cases where P-RFO fails from FSM guesses, the other multistep biradicaloid reaction etc.) 

\section*{References}
\bibliography{references}

\end{document}


\title
         {\large Supporting Information for Locating Ab Initio Transition States via Geodesic Construction on Machine Learned Potential Energy Surfaces}

	\author{Diptarka Hait}
        \author{Jan D. Estrada Pabón}
        \author{Martin St\"ohr}
        \author{Todd J. Mart{\'i}nez}
\email{todd.martinez@stanford.edu; toddjmartinez@gmail.com}
\affiliation
{{Department of Chemistry and The PULSE Institute, Stanford University, Stanford, California 94305, United States}}
\affiliation{SLAC National Accelerator Laboratory, Menlo Park, California 94024, United States}

\maketitle

\section{Comparison of P-RFO with Cartesian Coordinates to Delocalized Internal Coordinates for the FSM Development Dataset}
\begin{table}[h!]
\caption{Performance of P-RFO initialized with TS guess geometries from MLP geodesic and FSM. Values for optimization with Cartesian coordinates are provided in parentheses.}
\label{tab:fsmdevcart}
\centering
\begin{tabular}{l c c }
\toprule
\multirow{3}{*}{\textbf{Reaction}} & \multicolumn{2}{c}{\textbf{P-RFO iterations starting from}}  \\
 & \textbf{ MLP geodesic} & \textbf{FSM}  \\
  & \textbf{guess} & \textbf{guess}  \\
\midrule
\ce{H2CO -> H2 + CO} & 3 (3) & 8 (11) \\
\ce{SiH4 -> SiH2 + H2} & 3 (3) & 3 (3)  \\
\ce{CH3CHO -> H2C=CHOH} & 3 (3) & 4 (4) \\
2,4-hexadiene \ce{->} 3,4-dimethylcyclobutene & 10 (38) & 9 (33) \\
Alanine dipeptide rearrangement & 58 (fails) & 56 (fails)   \\
\parbox[t]{3in}{\raggedright Ireland–Claisen rearrangement between silyl ketene acetal and silyl ester} & \multirow{2}{*}{59 (not done)} & \multirow{2}{*}{69  (not done)} \\
\ce{CH3CH3 -> H2C=CH2 + H2}  & 4 (4) & fails (24)  \\
bicyclobutane \ce{-> H2C=CH-CH=CH2}  & 4 (4) & 39 (39)   \\
\ce{H2C=CH-CH=CH2 + H2C=CH2 ->} cyclohexene  & 6 (6) & 22 (60)  \\
\bottomrule
\end{tabular}
\end{table}

\newpage
\section{Reactions from EW-CI-NEB dataset where FSM fails}
\setlength{\tabcolsep}{9pt}
\begin{table}
\caption{Summary of TS optimizations from the EW-CI-NEB dataset reported in Ref.~\citenum{asgeirsson2021nudged} dataset where MLP geodesic guesses converge, but FSM fails in some form.}
\label{tab:ts_fails_fsm} 
\begin{tabular}{@{} L{0.31\textwidth} l L{0.24\textwidth} L{0.3\textwidth} @{}}
\toprule 
\textbf{System Name} & \textbf{ID} & \textbf{MLP Geodesic Guess} & \textbf{FSM Guess}\\
\midrule 
H exchange between \ce{NH2} and \ce{BH2} moieties of two \ce{H2B-NH2} & 4 & \textbf{9} iterations (internal) or \textbf{12} (Cartesian). & \textbf{Failed} to converge using either coordinate system. \\
\midrule
Hydroxyacetone formation from vinyl alcohol and formaldehyde & 14 & \textbf{16} iterations (internal) or \textbf{21} (Cartesian). & \textbf{Failed} with internal coordinates, \textbf{converged} with Cartesian in \textbf{45} iterations. \\
\midrule
2-propene-1,2-diol formation  from vinyl alcohol and formaldehyde & 19 & \textbf{6} iterations (internal) or \textbf{8} (Cartesian). & \textbf{Failed} with internal coordinates,  converged to an \textbf{incorrect TS} in \textbf{173} iterations with Cartesian. \\
\midrule
Vinyl alchohol and formaldehyde forming methanol and ketene& 22 & \textbf{10} iterations. &Converged to an \textbf{incorrect TS} in \textbf{185} iterations. \\
\midrule
Ethane dehydrogenation & 30 & \textbf{5} iterations (internal) or \textbf{4} (Cartesian). &\textbf{Failed} with internal coordinates, \textbf{converged} with Cartesian in \textbf{24} iterations. \\
\midrule
H exchange between methanol and formaldehyde & 54 & \textbf{3} iterations. & Converged to an \textbf{incorrect TS} in \textbf{169} iterations. \\
\midrule
H exchange between \ce{NH3} of \ce{H3B-NH3} and \ce{BH2} of \ce{H2B-NH2}  & 120 & \textbf{158} iterations (internal) or \textbf{191} (Cartesian). & \textbf{Failed} to converge using either coordinate system. \\
\bottomrule
\end{tabular}
\end{table}

\section{Multistep biradicaloid Reorganization between 1,3-butadiene and ethene}

\begin{figure}[htb!]
    \includegraphics[width=\linewidth]{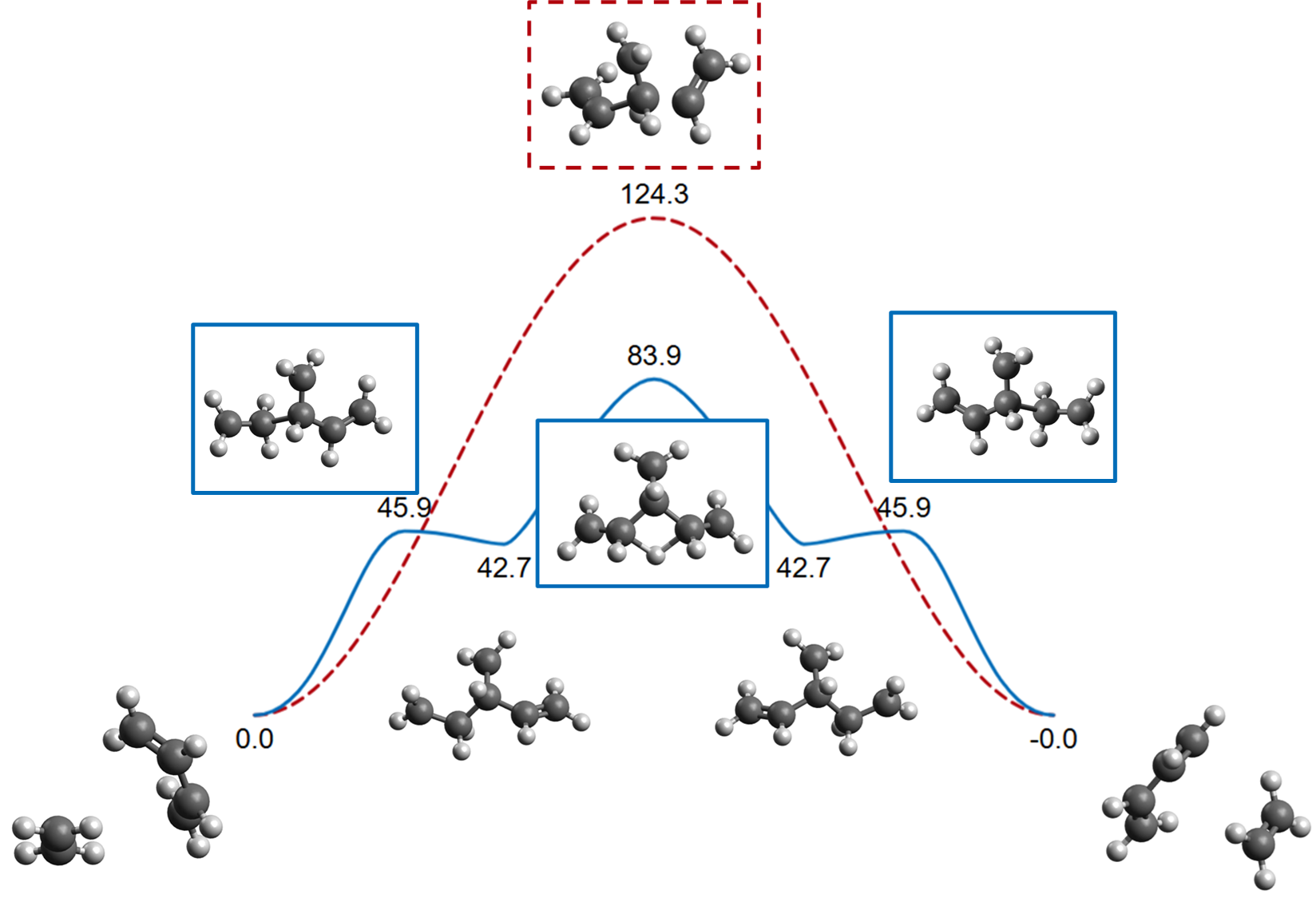}
    \centering
    \caption{The reorganization reaction between 1,3-butadiene and ethene from the Ref.~\citenum{asgeirsson2021nudged} dataset for which MLP geodesic construction identifies a lower energy multistep pathway (blue solid line) over a higher energy elementary process (red dotted line). Key stationary structures along each path are shown (with the transition states bordered by the line-style corresponding to the reaction mechanism), as are their energies (in kcal/mol, relative to the reactant). The formation of the biradicaloid intermediates involves addition of ethene to butadiene (barrier of 45.9 kcal/mol, as shown) and a conformer rearrangement involving rotation of the terminal \ce{CH2} about the former ethene \ce{CC} bond. The latter process is low energy and is thus not shown.}
    \label{fig:twosteps}
\end{figure}
\bibliography{references}